\documentclass{aa}  

\usepackage{graphicx}
\usepackage{subcaption}
\usepackage{txfonts}
\usepackage{url}      
\usepackage{bm}        
\usepackage{soul}
\usepackage{supertabular}
\usepackage{hyperref}
\usepackage{xcolor}
\newcommand{\msolar} {$\rm{M_{\odot}}$}
\newcommand{\msolarc} {$\rm{M_{\odot}}$}

\newcommand{\molH} {$\rm{H_2}$~}

\defcitealias{Prole2023}{LP23}
\begin{document}

\title{Heavy Black Hole Seed Formation in High-z Atomic Cooling Halos}
   \authorrunning{Lewis R. Prole}
   \titlerunning{Atomically cooled halos: massive black hole seeds}

   \author{Lewis R. Prole\inst{1}\fnmsep\thanks{E-mail: lewis.prole@mu.ie}
          John A. Regan\inst{1},
          Simon C. O. Glover\inst{2}, 
          Ralf~S.~Klessen\inst{2,3}, 
          Felix D. Priestley\inst{4}, 
          Paul C. Clark\inst{4}, 
          }

   \institute{\inst{1}Centre for Astrophysics and Space Science Maynooth, Department of Theoretical Physics, Maynooth University, Maynooth, Ireland.\\
   \inst{2}Universit\"{a}t Heidelberg, Zentrum f\"{u}r Astronomie, Institut f\"{u}r Theoretische Astrophysik, Albert-Ueberle-Stra{\ss}e 2, D-69120 Heidelberg, Germany.\\
   \inst{3}Universit\"{a}t Heidelberg, Interdisziplin\"{a}res Zentrum f\"{u}r Wissenschaftliches Rechnen, Im Neuenheimer Feld 205, D-69120 Heidelberg, Germany.\\
   \inst{4}Cardiff University School of Physics and Astronomy, United Kingdom.\\
             }

   \date{Received ; accepted }

 
  \abstract
   {Halos with masses in excess of the atomic limit are believed to be ideal environments in which to form heavy black hole seeds with masses above $10^3$ M$_\odot$. In cases where the \molH fraction is suppressed this is expected to lead to reduced fragmentation of the gas and the generation of a top heavy  initial mass function. In extreme cases this can result in the formation of massive black hole seeds. Resolving the initial fragmentation scale and the resulting protostellar masses has, until now, not been robustly tested.}
   {We run zoom-in simulations of atomically cooled halos in which the formation of \molH is suppressed to assess whether they can truly resist fragmentation at high densities and tilt the initial mass function towards a more top heavy form and the formation of massive black hole seeds.}
   {Cosmological simulations were performed with the moving mesh code {\sc Arepo} using a primordial chemistry network until $z \sim 11$. Three haloes with masses in excess of the atomic cooling mass were then selected for detailed examination via zoom-ins. A series of zoom-in simulations, with varying levels of maximum spatial resolution, capture the resulting fragmentation and formation of metal free stars using the sink particle technique. The highest resolution simulations resolve densities up to $10^{-6}$ g cm$^{-3}$ (10$^{18}$ cm$^{-3}$) and capture a further 100 yr of fragmentation behaviour at the center of the halo. Lower resolution simulations were then used to model the future accretion behaviour of the sinks over longer timescales.}
   {Our simulations show intense fragmentation in the central region of the halos, leading to a large number of near-solar mass protostars. Even in the presence of a  super-critical Lyman-Werner radiation field ($J_{LW} > 10^5 J_{21}$) \molH continues to form within the inner $\sim$2000 au of the halo. Despite the increased fragmentation the halos produce a protostellar mass spectrum that peaks at higher masses relative to standard Population III star forming halos. The most massive protostars have accretion rates of 10$^{-3}$-10$^{-1}$ M$_\odot$ yr$^{-1}$ after the first 100 years of evolution, while the  total mass of the central region grows at 1 M$_\odot$ yr$^{-1}$. Lower resolution zoom-ins show that the total mass of the system continues to accrete at $\sim$1 M$_{\odot}$ yr$^{-1}$ for at least 10$^4$ yr, although how this mass is distributed amongst the rapidly growing number of protostars is unclear. However, assuming that a fraction of stars can continue to accrete rapidly the formation of a sub-population of stars with masses in excess of $10^3$ \msolar \ is likely in these halos. In the most optimistic case we predict the formation of heavy black hole seeds with masses in excess of $10^4$ \msolar \ assuming an accretion behaviour in line with expectations from super-competitive accretion and/or frequent mergers with secondary protostars.}
   {}


   \maketitle


\section{Introduction} \label{sec:Introduction}

Quasi-stellar radio sources (quasars) are thought to be powered by supermassive black holes (SMBHs) which populate the centers of most, if not all, massive galaxies. Quasars have been detected out to redshifts of $z>7$ with black hole (BH) masses believed to be in excess of $10^{9}$ M$_\odot$ \citep[e.g.][]{Mortlock2011,Matsuoka2019}, implying that the seeds for these SMBHs formed in the early Universe. The existence of SMBHs with masses in excess of $10^9$ M$_\odot$ within the first billion years of the Universe poses a challenge to our understanding of both BH formation and BH accretion. How could such massive objects appear so early in cosmic history? Two mainstream pathways have emerged over the last four decades; SMBHs may originate from so-called light seeds with masses less than 10$^3$ \msolar, or from heavy BH seeds with masses significantly in excess of 10$^3$ \msolar.

Light seeds are typically thought to form from the remnants of the first stars - Population III (Pop III) stars. Initial ab-initio modelling of Pop III star formation predicted an extremely top heavy IMF for the first stars \citep[e.g.][]{Bromm1999, Bromm2001, Abel2002a} with characteristic masses of order 1000 \msolarc. However, more recent studies throughout the last decade have found that Pop III stellar masses are lower than initially suggested (e.g. \citealt{Stacy2010b,Clark2011a,Hirano2014,Susa2019,Prole2023}; for a recent review see \citealt{Klessen2023}) with characteristic masses of a few tens of M$_\odot$. Depending on the exact onset of BH formation these light seeds would need to maintain accretion at the Eddington rate for several hundred million years in order to bridge the approximately seven orders of magnitude in mass required to reach the upper limits of the SMBH threshold. Alternatively periods of super-Eddington accretion may offer a solution. In this case the light seed BH can grow at supra-exponential rates \citep{Alexander2014} which only need to last for brief periods of time \citep{Lupi2016, Inayoshi2016}. However, in both Eddington limited and super-Eddington cases, radiative feedback from the accretion of matter onto BHs heats the surrounding gas and lowers the accretion rate \citep{Johnson2007,Milosavljevic2009,Alvarez2009, Smith2018} - making maintaining either Eddington accretion and/or super-Eddington accretion unlikely over sustained periods \citep{Regan2019, Su2023, Massonneau2023}. Finally, both Eddington and super-Eddington limited growth requires that the light seed sit at the centre of a powerful gas inflow and that the embryonic BH can readily accrete the surrounding dense gas. However, dynamical studies of BHs have shown that this is also a challenge, with light seeds tending to walk random trajectories around the host halo centres \citep{Beckmann2019, Pfister2019}. As a result of these obstacles to light seed growth, the possibility of much heavier BH seeds has also been studied, beginning with \cite{Rees1978}.

Forming heavy seeds has hinged on two separate but not necessarily distinct pathways. On the one hand dynamical processes have been invoked to explain heavy BH seed formation; runaway collisions in dense young star clusters could produce massive black holes (MBHs) of $\sim 10^3$ M$_\odot$  \citep[e.g.][]{PortegiesZwart2004, Glebbeek2009, Katz2015,Reinoso2023} that are candidates for the ultraluminous X-ray sources observed in young star forming regions  \citep{Ptak2004}. These MBHs may grow into SMBHs through binary mergers and/or gas accretion \citep{Micic2007}. Additionally, BH mergers within a dense BH cluster may achieve the same outcome \citep{Stone2017, Schleicher2022} through either collisions and/or gas accretion, although the growth prospects are far from certain \citep{ArcaSedda2023}. \\
\indent On the other hand conditions may exist in the early Universe conducive to the formation of truly massive stars with final masses well in excess of $10^3$ \msolar \ \citep{Regan2020, Latif2022, Regan2023} and possibly up as high as a few times $10^5$ \msolarc \citep{Woods2017}. The formation of these massive primordial Pop III stars requires high inflow rates onto the stellar surface \citep{Haemmerle2018, Woods2020} but repeated numerical
experiments have shown the stars to be stable for at least 2 Myr until their inevitable direct collapse into a MBH \citep[e.g.][]{Hosokawa2009,Hosokawa2012a}. Finally, it may also be that the processes that lead to a dense stellar cluster or a massive Pop III star form part of a continuum with 
a massive star forming at the very centre of a dense stellar cluster where collisions drive the formation of a very massive star \citep{Boekholt2018, Chon2020, Schleicher2023}.
The goal of this paper will be to test whether so-called atomic cooling halos can lead to the formation of a heavy seed Pop III star using state-of-the-art, high resolution, hydrodynamic simulations.

Pristine atomic cooling halos have long been suggested as the ideal environment in which to seed MBHs \citep{Loeb1994, Spaans2006, Prieto2013}. If the \molH abundance inside the massive halo can be suppressed then the gas must cool predominatly through atomic hydrogen line emission and H$^{-}$ free-bound emission. 
The thermal pathway then taken by the gas inside an atomic cooling halo in the absence of effective \molH cooling deviates significantly from the standard Pop III star formation scenario (e.g. \citealt{Omukai2000,Klessen2023}). \\
\indent Pop III star formation in minihalos is facilitated by molecular hydrogen. The DM halo potential well pulls in the gas and shock heats it up to $\sim 1000$ K, At these temperatures, the H$_2$ abundance increases to $\sim 10^{-4}$ (e.g. \citealt{Tegmark1997,Greif2008}) where rotational transitions can occur via electrical quadrupole radiation, which allows the gas to cool down to minimum temperatures of $\sim$200 K (e.g. \citealt{Abel1997,Bromm1999,Glover2008}) and collapse, decoupling from the DM halo. \\
\indent One potential way in which H$_2$ abundances can be reduced is via a nearby source of Lyman-Werner (LW) radiation. These far-ultraviolet (FUV) photons in the Lyman and Werner bands of H$_{2}$ (11.2 -- 13.6~eV) can dissociate H$_2$ via the two-step Solomon process \citep{Field1966,Stecher1967}. Additionally, photons of above 0.76eV can photodissociate H$^{-}$, disrupting the primary H$_{2}$ formation channel (e.g. \citealt{Chuzhoy2007}). Before the Str{\"o}mgren spheres of Pop III stars overlap, the UV background below the ionization threshold is able to penetrate large clouds and suppress their H$_2$ abundance \citep{Haiman1997}. This photodissociation of H$_2$ suppresses further star formation inside small halos and delays reionization until larger halos form \citep{Haiman2000}. While other physical mechanism can have a similar effect, we are focused here in studying the gas collapse inside atomically cooling halos, which are
both pristine and have had their \molH cooling efficiency suppressed.


In halos with virial temperatures below $T_{\rm vir} \sim 8000$~K, star formation is suppressed entirely if LW radiation reduces the H$_{2}$ abundance below the level at which gas can cool within a Hubble time. For example, an intense burst of LW radiation from a neighbouring star-bursting protogalaxy just before the gas cloud undergoes gravitational collapse is proposed to prevent the cloud from collapsing or forming stars \citep{Regan2017}. However, the halo will continue to grow through hierarchical mergers (e.g. \citealt{Chon2016,Dong2022}). Once the virial temperature reaches $\sim 8000$~K, it becomes possible for the gas to cool via Lyman-$\alpha$ emission. The required virial temperature is related to the virial mass through the relation \citep{Fernandez2014}
\begin{equation}
\rm T_{vir} = 0.75 \times 1800 \Big(\frac{M}{10^6 M_{\odot}}\Big)^{2/3} \Big(\frac{1 + z}{21}  \Big),
\end{equation}
giving a virial mass of $\sim 3 \times 10^7$ M$_\odot$ at $z = 12$ for a virial temperature of 8000 K. If a halo can grow to this mass it will become hot enough to cool via atomic line emission and begin to collapse  \citep{Oh2002,Bromm2003,Bromm2011}. In this scenario, the collapse occurs almost isothermally, and fragmentation of the gas is thought to be suppressed throughout. 

The maximum density that can be reached in simulations of this process is related to the resolution of the simulation. To avoid artificial fragmentation, it is necessary to resolve the Jeans length, which progressively shrinks as the gas collapses to higher densities \citep{Truelove1997}. Therefore, the better the resolution of the simulation, the higher the density that it can reach. We know from simulations of the standard Pop III star formation scenario that the formation of the primordial protostar occurs at densities of 10$^{-6}$ - 10$^{-4}$ g cm$^{-3}$ where the gas becomes adiabatic \citep{Omukai2000,Machida2015}. The Jeans length at this stage is 0.01-0.1 au, and so the required resolution is roughly a factor of ten smaller than this. Despite this, the resolution of most atomically cooled halo simulations is relatively poor in comparison, with most studies not resolving their gas past densities of 10$^{-17}$ g cm$^{-3}$ (e.g. \citealt{Shang2010,Sugimura2014,Regan2014,Hartwig2015d,Agarwal2015,Glover2015,Agarwal2016,Regan2017,Dunn2018}). This resolution is sufficient to determine whether or not H$_{2}$ cooling is important during the initial collapse of the gas, but does not allow one to draw conclusions about the later stages of the collapse. Some studies find evidence for small-scale fragmentation, even in the absence of effective H$_{2}$ cooling (e.g. \citealt{Becerra2015,Becerra2018,Chon2018,Latif2020,Patrick2023}). The number of fragments formed is generally much smaller than the number found in recent simulations of the standard Pop III star formation scenario. However, this may be a consequence of the limited peak density: in most of these studies, the gas density never exceeds $\rho \sim 10^{-10} \: {\rm g \: cm^{-3}}$, four orders of magnitude smaller than the point at which we expect the collapse to become adiabatic \citep{Becerra2018a}.

This study aims to provide the most accurate picture of atomically cooled halo collapse at high densities to date, answering whether atomic cooling halos do experience reduced fragmentation and higher stellar/BH seed masses compared to the Pop III minihalo scenario, or whether the fragmentation at high densities produces protostellar masses  similar to what we have seen in simulations of Pop III star forming minihalos. To that end, we simulate the collapse of atomically cooled halos from cosmological initial conditions with zoom-in simulations running up to the lower limit of protostellar formation at 10$^{-6}$ g cm$^{-3}$ and capturing a further $\sim$100 yr of disc fragmentation. We directly compare our results to a recent study, \cite{Prole2023} (hereafter \citetalias{Prole2023}) which examined H$_2$ cooling minihalos with the same simulation code, chemical set-up and maximum resolution as the simulations presented in this work.

The format of this paper is as follows: in \S \ref{sec:method} we outline the numerical technique used including the simulation code and chemical network. In \S \ref{sec:collapse} we discuss the collapse of the gas up to point immediately prior to sink formation, while in \S \ref{sec:frag} we discuss the fragmentation of the gas after the insertion of sink particles. In \S \ref{sec:mass} we analyse the growth of the stellar system and compare, via a convergence study, the evolution of the star particles into main sequence stars and discuss their eventual evolution into (massive) BHs. In \S \ref{sec:caveats} we discuss some caveats before concluding in \S \ref{sec:conclusions}.

\section{Numerical method}
\label{sec:method}
\subsection{{\sc Arepo}}
The simulations presented here were performed with the moving mesh code {\sc Arepo} \citep{Springel2010} with a primordial chemistry set-up described in \S \ref{sec:chem}. {\sc Arepo} combines the advantages of adaptive mesh refinement (AMR: \citealt{Berger1989}) and smoothed particle hydrodynamics (SPH: \citealt{Monaghan1992}) with a mesh made up of a moving, unstructured, Voronoi tessellation of discrete points. {\sc Arepo} solves hyperbolic conservation laws of ideal hydrodynamics with a finite volume approach, based on a second-order unsplit Godunov scheme with an exact Riemann solver. Automatic and continuous refinement overcome the challenge of structure growth associated with AMR (e.g. \citealt{Heitmann2008}). 



\subsection{Chemistry}
\label{sec:chem}
Collapse of primordial gas is closely linked to the chemistry involved (e.g.\ \citealt{Glover2006,Yoshida2007, Glover2008, Turk2011}). We therefore use a fully time-dependent chemical network to model the gas. We use the treatment of primordial chemistry and cooling originally described in \cite{Clark2011}, but with updated values for some of the rate coefficients, as summarised in \cite{Schauer2019}. The network has 45 chemical reactions to model primordial gas made up of 12 species: H, H$^{+}$, H$^{-}$, H$^{+}_{2}$ , H$_{2}$, He, He$^{+}$, He$^{++}$, D, D$^{+}$, HD and free electrons. Optically thin H$_{2}$ cooling is modelled as described in \citet{Glover2008}: we first calculate the rates in the low density ($n \rightarrow 0$) and LTE limits, and the smoothly interpolate between them as a function of $n / n_{\rm cr}$, where $n_{\rm cr}$ is the H$_{2}$ critical number density above which collisions are so frequent that they keep the populations close to their LTE values. To compute the H$_{2}$ cooling rate in the low density limit, we account for the collisions with H, H$_{2}$, He, H$^{+}$ and electrons. To calculate the H$_{2}$ cooling rate in the optically thick limit, we use an approach based on the Sobolev approximation \citep{Yoshida2006, Clark2011}. Prior to the simulation, we compute a grid of optically thick H$_{2}$ cooling rates as a function of the gas temperature and H$_{2}$ column density. During the simulation, if the gas is dense enough for the H$_{2}$ cooling to potentially be in the optically thick regime ($\rho > 2 \times 10^{-16} \: {\rm g \: cm^{-3}}$), we interpolate the H$_2$ cooling rate from this table, using the local gas temperature and an estimate of the effective H$_{2}$ column density computed using the Sobolev approximation. In addition to H$_{2}$ cooling, we also account for several other heating and cooling processes: cooling from atomic hydrogen and helium, collisionally-induced H$_{2}$ emission, HD cooling, ionisation and recombination, heating and cooling from changes in the chemical make-up of the gas and from shocks, compression and expansion of the gas, three-body H$_{2}$ formation and heating from accretion luminosity. For reasons of computational efficiency, the network switches off tracking of deuterium chemistry\footnote{Note that HD cooling continues to be included in the model.} at densities above 10$^{-16}$~g~cm$^{-3}$, instead assuming that the ratio of HD to H$_{2}$ at these densities is given by the cosmological D to H ratio of 2.6$ \times $10$^{-5}$. Note that although our treatment of H$_2$ cooling accounts for the opacity of the gas at high densities, our treatment of the effects of other cooling processes, such as H$^{-}$ free-bound emission, does not currently account for the continuum opacity of the gas. At densities below $\sim 10^{-8} \: {\rm g \, cm^{-3}}$, this makes little difference to the thermal evolution of the gas, but it means that we will tend to overestimate the cooling rate at densities above this value. The adiabatic index of the gas is computed as a function of chemical composition and temperature with the {\sc Arepo} HLLD Riemann solver. 


\subsection{Simulation Setup}
As discussed in the Introduction the goal of this study is to investigate the formation of primordial Pop III stars at the centre of a metal-free
atomically cooling halo. Such halos have previously been investigated as ideal sites for heavy seed formation \citep[][]{Haiman1996,Regan2009, Latif2013b,Latif2015a,Latif2015,Latif2020,Wise2019}. To generate the appropriate initial conditions we generated cosmological initial conditions using MUSIC \citep{Hahn2011}. An initial cosmological simulation was performed within a co-moving box of side length $1 \: h^{-1} \: {\rm Mpc}$ using a $\Lambda$CDM cosmology with parameters $h=0.6774$, $\Omega_0 = 0.3089$, $\Omega_{\rm b} = 0.04864$, $\Omega_\Lambda = 0.6911$, $n = 0.96$ and
$\sigma_8 = 0.8159$ \citep{Planck-Collaboration2020}.  The simulations were initialized at $z=127$ with an initial dark matter distribution using the transfer functions of \cite{Eisenstein1998}. The gas distribution was set within {\sc Arepo} to initially follow the dark matter (i.e. GENERATE\_GAS\_IN\_ICS = 1). We model the dark matter with $512^{3}$ particles and the gas was modelled with $512^{3}$ grid cells (prior to refinement). During the simulation, an additional Jeans refinement criterion was applied such that the Jeans length of the gas is always resolved with at least 4 grid cells. Hence, the gas is able to dynamically refine during the simulation allowing maximum resolution where required. \\
\indent During this initial phase, we disabled the molecular chemistry functionality and hence utilised a simpler 6 species model such that H$_2$ and HD abundances remained at their initial value. This eliminated the need for a LW background radiation field at this initial stage of the calculation since the goal of this study is to investigate the idealised case of Pop III formation inside of a pristine, atomically cooling halo at a virial temperature of approximately 8000 K. We note that applying a super-critical LW field has the same result on the H$_2$ abundance. As the dominant HD formation pathway relies on the H$_2$ abundance \citep{Nakamura2002}, a super-critical LW field also suppresses the HD abundance.

\begin{figure*}
\centering
\begin{subfigure}[b]{0.55\textwidth}
	 \hbox{\hspace{-1cm} \includegraphics[scale=0.62]{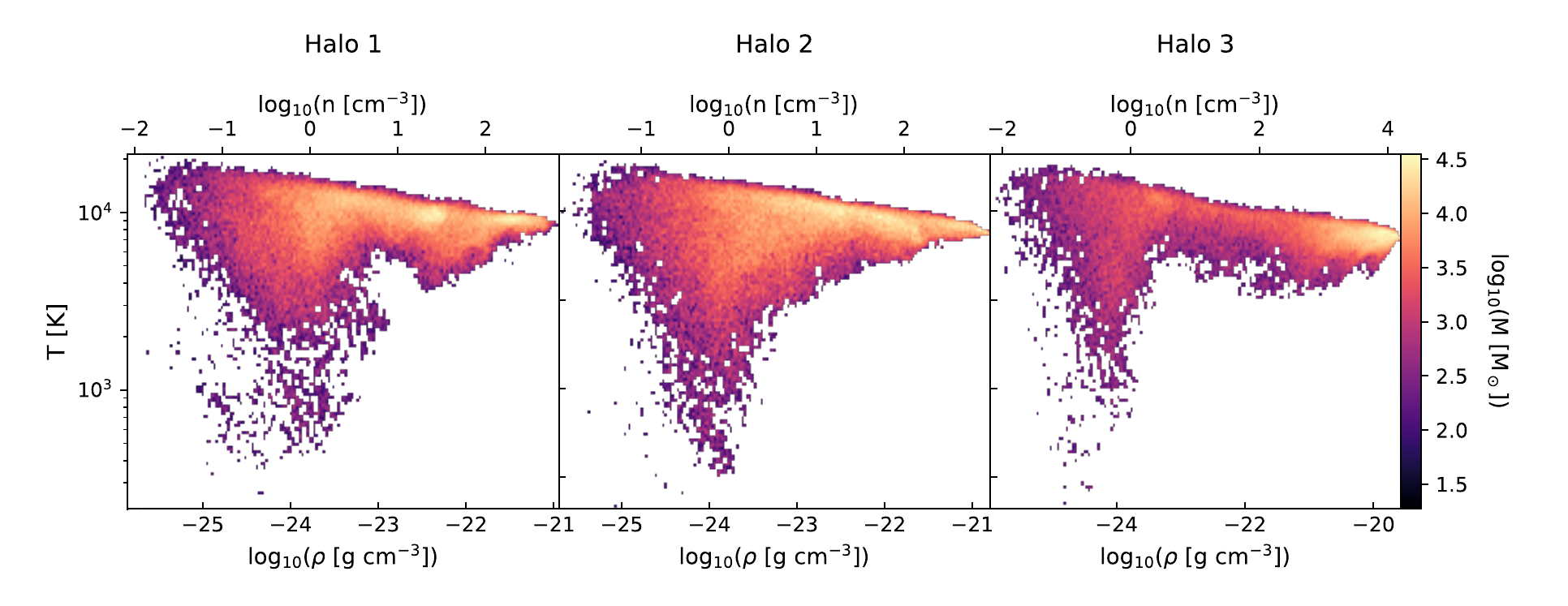}}
\end{subfigure}
    \caption{2D histograms of the temperature - density profiles for the halos at the point at which they are extracted from the initial cosmological simulation, weighted by total gas mass within each 2D bin. The halos have begun to gravitationally collapse via atomic cooling as seen from the close to isothermal temperature profile. They differ from the regular H$_2$ minihalo case by the absence of a sharp drop in temperature to $\sim$200 K beginning at $\sim 10^{-23}$ g cm$^{-3}$.}
    \label{fig:ics1}
\end{figure*}

\begin{figure*}
\centering
\begin{subfigure}[b]{0.55\textwidth}
	 \hbox{\hspace{-1cm} \includegraphics[scale=0.62]{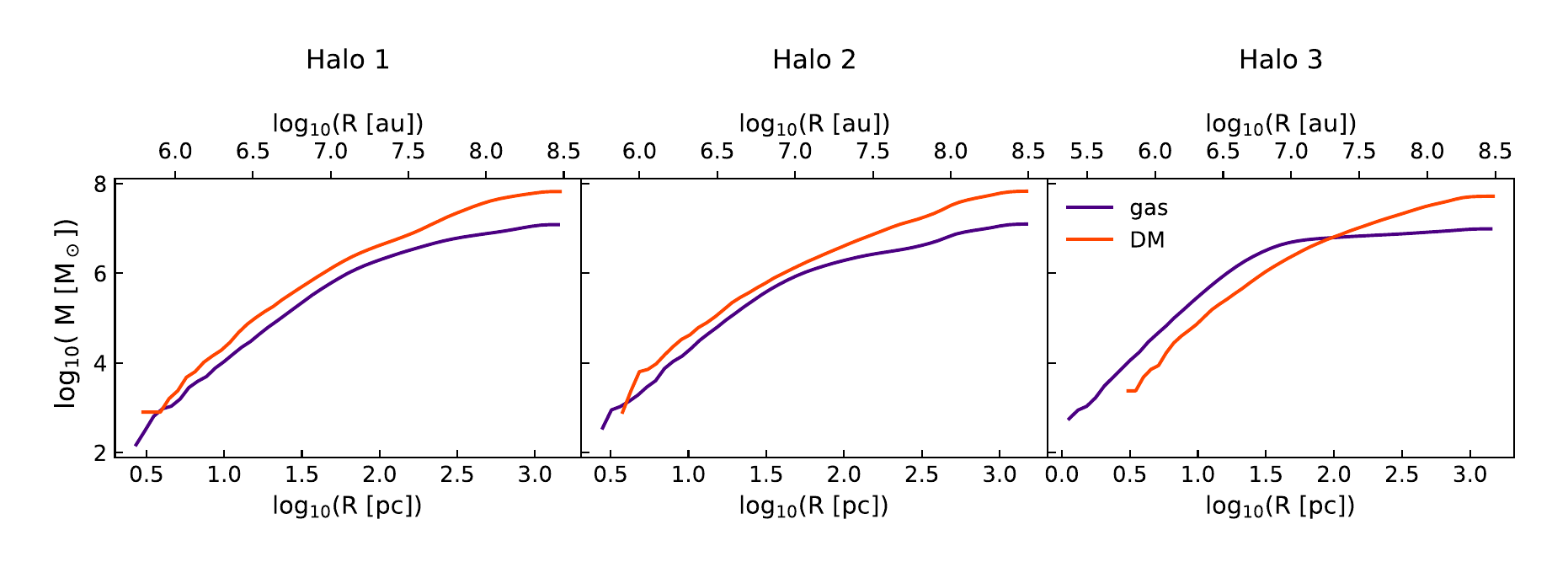}}
\end{subfigure}
    \caption{Radial profiles of enclosed gas and DM mass for the three atomic halos at the point where they are extracted from the initial cosmological simulation. The baryonic component becomes comparable to the DM within the central $\sim100$ pc where it begins to decouple from the DM. In Halo 3, the baryonic component is dominant over DM on these scales.}
    \label{fig:ics2}
\end{figure*}

At $z \sim 13.3$ the first atomic halo begins to collapse within our box. Two more, physically distinct halos collapse at z = 11.5. These first three halos to begin atomically cooling and collapsing were then extracted. The selection criteria was simple; while H$_2$ rich halos begin cooling down from 1000 K from $\sim$10$^{-23}$ g cm$^{-3}$, reaching 200 K by $\sim$10$^{-22}$ g cm$^{-3}$,  we have disabled H$_2$ formation, hence the only way for the gas to collapse to these densities is via atomic cooling. A density threshold for selecting a collapsing halo was therefore chosen as slightly higher than this density at 10$^{-21}$ g cm$^{-3}$. Only halos containing gas cells with a density exceeding this threshold were selected for extraction and resimulation. The central coordinate of the halos was found using a friends of friends (FoF) algorithm and a new box length of 2 kpc (physical units) was cut around it. Using the new box cut from the parent box, the simulations were restarted as "zoom-in" simulations. The units were converted into physical units for the remainder of the zoom-in calculation and the simulation essentially run as an isolated galaxy simulation with periodic boundary conditions. Table \ref{table:1} shows the virial mass, radius and temperature of each halo as calculated by the FoF algorithm as well as the redshift when it was extracted, while Figure \ref{fig:ics1} shows the temperature-density profiles at that stage and Figure \ref{fig:ics2} shows the radial profiles of the enclosed mass for both the gas and the DM.

\begin{figure*}
\centering
\begin{subfigure}[b]{0.55\textwidth}
	 \hbox{\hspace{-2.2cm} \includegraphics[scale=0.7]{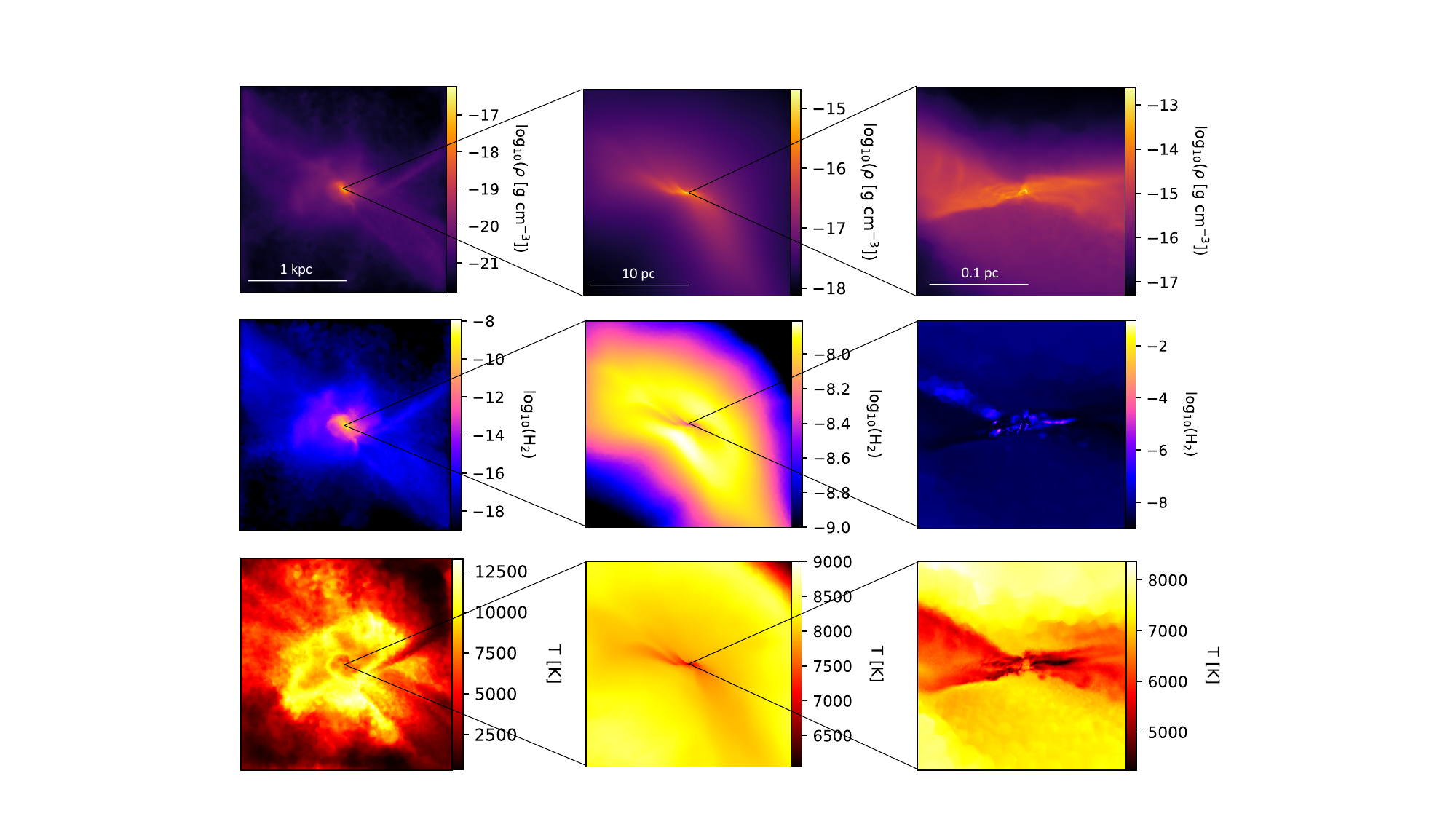}}
\end{subfigure}
    \caption{Projection images of density, H$_2$ fraction and temperature for Halo 3 at the end of the simulation. From left to right, the zoom-in plots show at radius of 1 kpc, 10 pc and 0.1 pc, respectively.}
    \label{fig:mosaic}
\end{figure*}

For the zoom-in simulations, the full chemistry model (12 species) is again enabled to get an accurate picture of \molH formation at very high densities inside the core of the collapsing halo. We invoke a super-critical LW radiation field of $J_{LW} = 10^5 J_{21}$ (where $J_{21}$ is in units of 10$^{-21}$~erg~s$^{-1}$~cm$^{-2}$~Hz$^{-1}$~sr$^{-1}$) to suppress H$_2$ formation. We model the effects of H$_2$ self-shielding using the TreeCol algorithm \citep{Clark2012}. \\
 \indent We turn-on the 12 species model to examine the impact of \molH production which is inevitable at the highest densities. In our highest resolution simulations, we evolve the collapse up to a density of 10$^{-6}$ g cm$^{-3}$ before inserting sink particles (see \S \ref{sec:sinks}) and capturing a further $\sim$100 yr of disc fragmentation and accretion behaviour, while in the lowest resolution simulation we capture $\sim$10$^4$ yr of accretion. See Table \ref{table:2} for a list of simulation realisations and resolutions employed. The structure of Halo 3 at the end of the high resolution simulation is visualised in Figure \ref{fig:mosaic}, which shows the density field along with H$_2$ fraction and temperatures at various zoom-in scales.

\begin{figure*}
\centering
\begin{subfigure}[b]{0.55\textwidth}
	 \hbox{\hspace{-1.3cm} \includegraphics[scale=0.70]{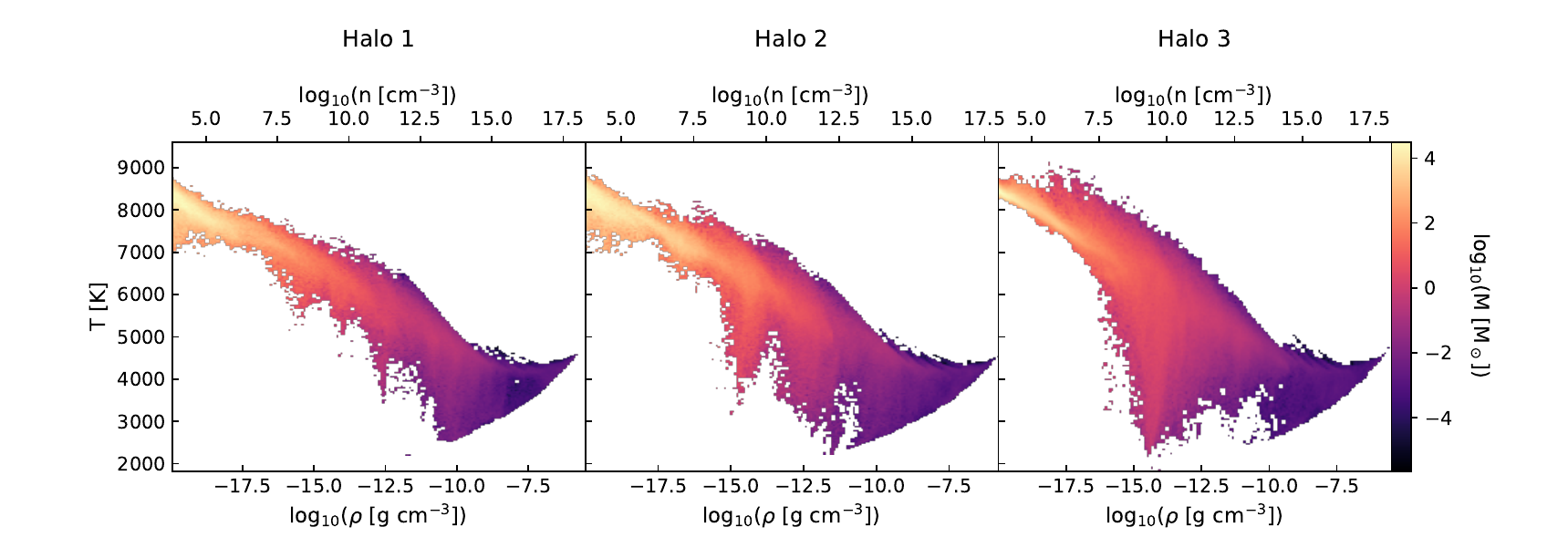}}
\end{subfigure}

 \ \ 
 
\begin{subfigure}[b]{0.8\textwidth}
	 \hbox{\hspace{-1.3cm} \includegraphics[scale=0.70]{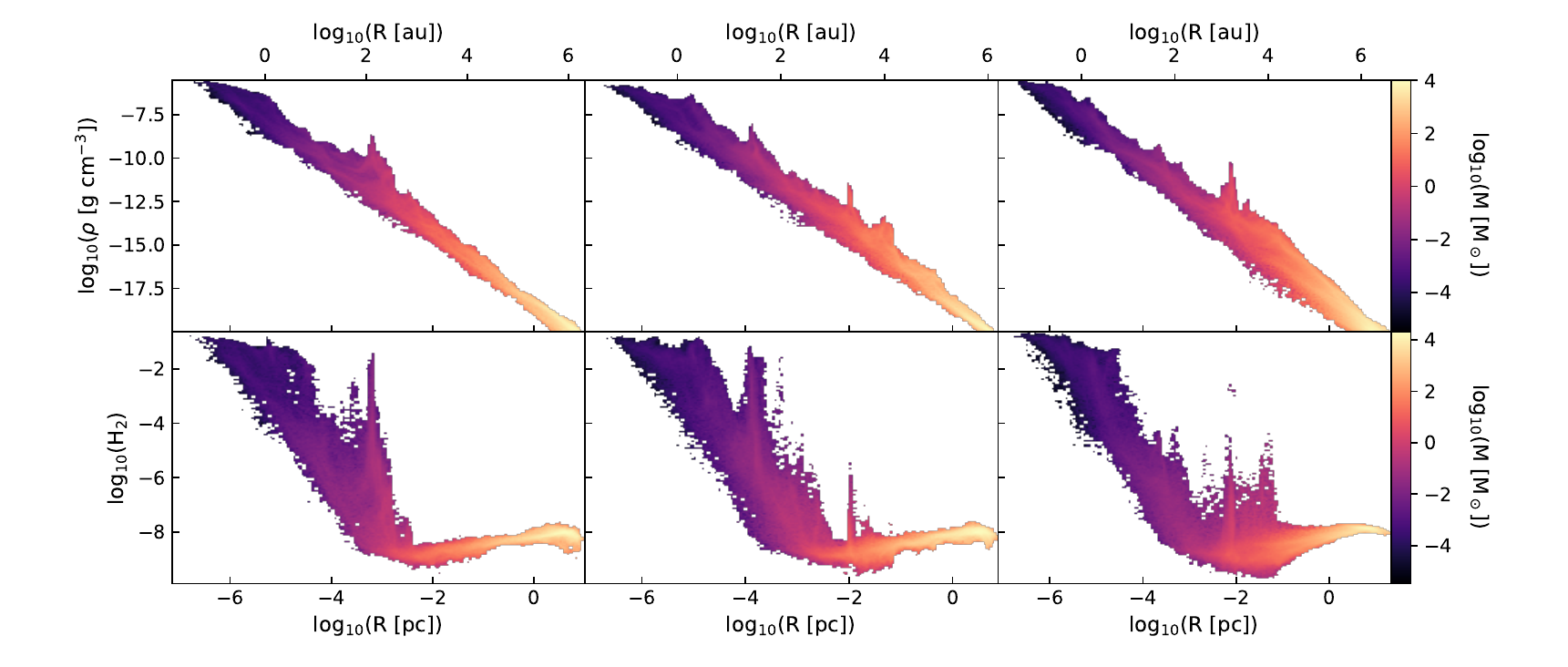}}
\end{subfigure}
    \caption{2D histograms of the characteristics of the halos at the point just before the formation of the first sink particle i.e. when the simulations approach their maximum density, weighted by total gas mass within each 2D bin. \textit{Top Panel}: Temperature-density relation. The collapse is close to isothermal at approximately 8000 K, with the temperature decreasing by only a factor of two over more than ten orders of magnitude in density. \textit{Bottom Panel:} Radial profiles of density and H$_2$ abundance. The density-radius relationship follows the $\rho \propto R^{-2}$ relationship expected for an isothermal collapse. The \molH abundances are initially negligible at the halo virial radius (R $\sim$ 100 pc) but increases in the centre of the halo once the density becomes high enough for self-shielding and three-body H$_2$ formation to become effective).}
    \label{fig:sink_formation}
\end{figure*}

\begin{figure}
	 \hbox{\hspace{-1cm} \includegraphics[scale=0.75]{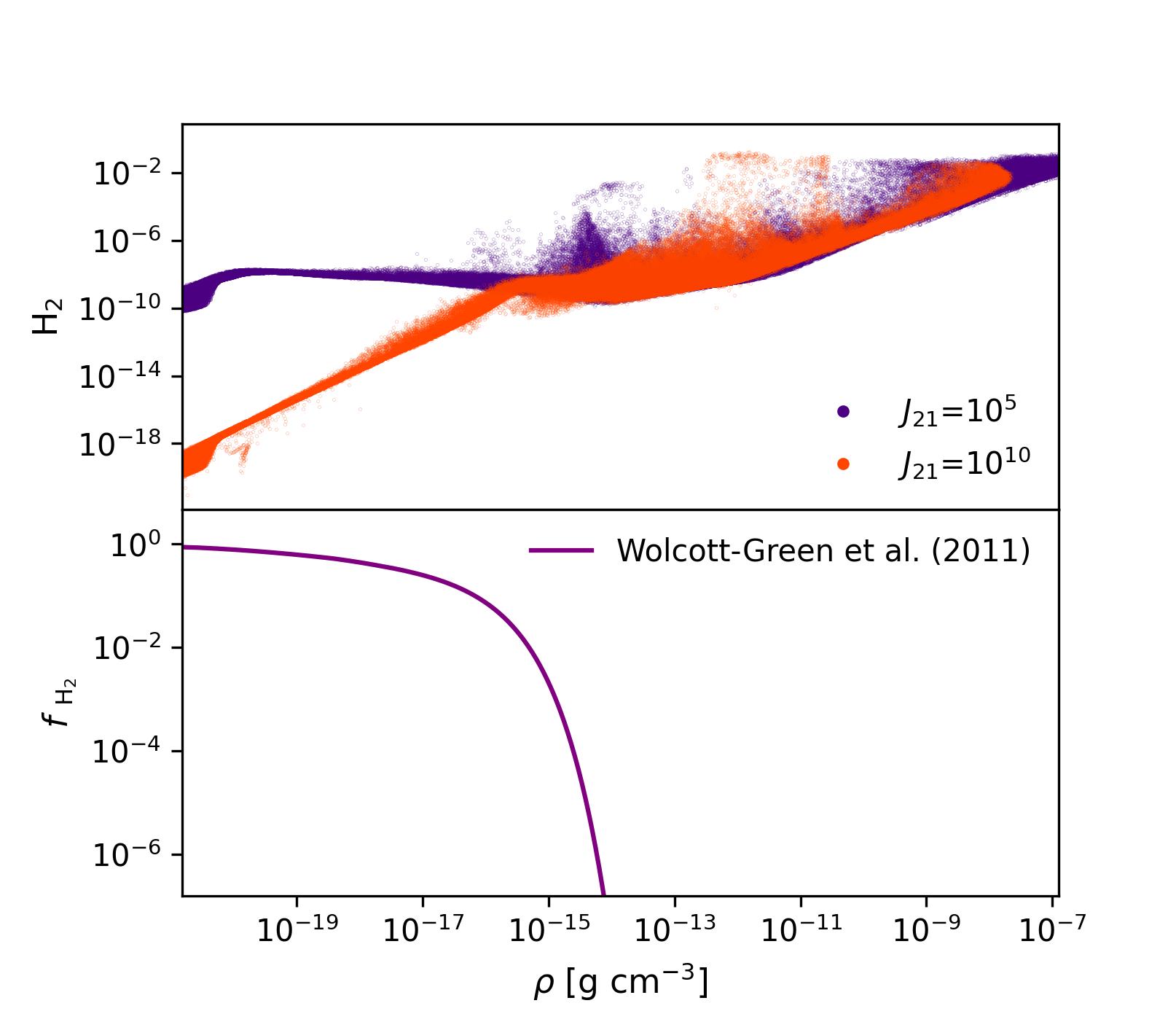}}
    \caption{Top - Comparison of the build up of H$_2$ in high density regions when a $J_{21}=10^5$ and 10$^{10}$ LW radiation field are used. Bottom - H$_2$ shielding factor calculated using equation 12 from \citet{Wolcott-Green2011}.}
    \label{fig:shielding}
\end{figure}

\begin{table}
	\centering
	\caption{Halo virial masses, radii and temperatures along with redshifts at the point where they are extracted from the initial cosmological simulation.}
	\label{table:1}
	\begin{tabular}{ccccc} 
		\hline
		Halo & M$_{200}$ [M$_{\odot}$] & R$_{200}$ [kpc] & T$_{\rm vir}$ [K] & $z$ \\
		\hline
		1 & 6.89 $\times$ 10$^7$ & 1.02 & 13536.60 & 11.5  \\
		2 & 5.86 $\times$ 10$^7$ & 0.85 & 13890.77 & 13.3   \\
		3 & 5.61 $\times$ 10$^7$ & 0.95 & 11794.74 & 11.5   \\
		\hline
	\end{tabular}
\end{table}


\subsection{Low resolution re-simulations}
As our high resolution simulations are only able to evolve for approximately one hundred years after the formation of the first protostar due to their extreme computational cost, it is not possible to directly determine the zero age main sequence mass of the stars formed in the system. We therefore re-run the simulation of Halo 3 with lower resolution to make an estimate of protostellar growth on longer timescales. The resolution study presented in \cite{Prole2022} found that the total mass accreted across all sink particles is well estimated by low resolution simulations, with less fragmentation yielding higher mass protostars. Our lower resolution simulations can therefore estimate the total mass in sink particles long after the high resolution simulations end. Since the three atomically cooled halos examined here have a similar total sink particle mass evolution (see \S \ref{sec:frag}), we only re-ran lower resolution simulations of Halo 3, which should nonetheless be a good proxy for all three halos. \\
\indent We chose to re-run the simulation with threshold sink particle creation densities of 10$^{-13}$ and 10$^{-10}$ g cm$^{-3}$, changing the minimum cell volume and gravitational softening lengths appropriately (see \S \ref{sec:sinks} for details). The three different density threshold simulations will be referred to as high, medium and low resolution from here on, although we emphasise that even the low resolution simulations have an extremely high spatial resolution of $\sim$300 au.


\begin{table*}
	\centering
	\caption{Simulation parameters. Left to right - sink particle creation density, accretion radius, gas minimum cell volume and minimum gravitational softening length, DM mass resolution and gravitational softening length.}
	\label{table:2}
	\begin{tabular}{ccccccc} 
		\hline
		resolution & $\rho_{\rm sink}$ [g cm$^{-3}$]& R$_{\rm sink}$ [au] & V$_{\rm min}$ [au$^3$] & L$_{\rm soft}$ [au] & M$_{\rm DM}$ [M$_\odot$] & L$_{\rm DM}$ [pc]\\
		\hline
		low & 10$^{-13}$ & 264.11 & 2.88$\times$ 10$^6$ & 132.02 & 793.25 & 1.18\\
		medium & 10$^{-10}$ & 9.20  & 12.18 &  4.60  & 793.25 & 1.18 \\
		high & 10$^{-6}$ & 0.11  & 2.33$\times$ 10$^{-5}$ & 0.057 & 793.25 & 1.18 \\
		\hline
	\end{tabular}
\end{table*}

\subsection{Sink particles}
\label{sec:sinks}
The simulation mesh must be refined during a gravitational collapse to ensure the local Jeans length is resolved. Once the mesh refines down to its minimum cell volume, sink particles must be introduced to represent the dense gas, preventing artificial instability in cells larger than their Jeans length. Our sink particle implementation was introduced in \cite{Wollenberg2020} and \citet{Tress2020}. A cell is converted into a sink particle if it satisfies three criteria:

   \begin{enumerate}
      \item The cell reaches a threshold density.
      \item It is sufficiently far away from pre-existing sink particles so that their accretion radii do not overlap.
      \item The gas occupying the region inside the sink is gravitationally bound and collapsing.
   \end{enumerate}
   Likewise, for the sink particle to accrete mass from surrounding cells, the cell must meet two criteria: 
      \begin{enumerate}
      \item The cell lies within the accretion radius.
      \item It is gravitationally bound to the sink particle.
   \end{enumerate}
   A sink particle can accrete up to 90$\%$ of a cell's mass, above which the cell is removed and the total cell mass is transferred to the sink. 

Increasing the threshold density for sink particle creation drastically increases the degree of fragmentation, reducing the masses of subsequent secondary protostars \citep{Prole2022}. However, increasing the sink particle threshold density also increases the computational challenge beyond which it is currently intractable. Ideally, sink particles would be introduced when the gas becomes adiabatic at $\sim 10^{-4}$ g cm$^{-3}$ \citep{Omukai2000}. However, the zero metallicity protostellar model of \cite{Machida2015} suggests that stellar feedback kicks in to halt collapse at $\sim 10^{-6}$ g cm$^{-3}$ (10$^{18}$ cm$^{-3}$), so we choose this as our sink particle creation density for our highest density simulations.

The accretion radius of a sink particle $R_{\text{sink}}$ is chosen to be the Jeans length $\lambda_{\text{J}}$ corresponding to the sink particle creation density and corresponding temperature. Taking our high resolution simulation as an example, at a density of $10^{-6}$ g cm$^{-3}$, we take the temperature value from \citet{Prole2022} of 4460 K to give a Jeans length of $1.67 \times 10^{12}$ cm (0.11 au). We set the minimum cell length to fit 8 cells across the sink particle accretion radius in compliance with the Truelove condition, by setting a minimum cell volume $V_{\text{min}}=(R_{\text{sink}}/4)^3$. The minimum gravitational softening length for cells and sink particles $L_{\text{soft}}$ is set to $R_{\text{sink}}/2$. The simulation parameters for the low, medium and high resolution simulations are summarised in Table \ref{table:2}.

The sink particle treatment also includes the accretion luminosity feedback from \cite{Smith2011}, as implemented in {\sc Arepo} by \citet{Wollenberg2020}. Stellar internal luminosity is not included in this work, which is not a problem in our high resolution simulations because the Kelvin-Helmholtz times of the protostars formed are much longer than the period simulated, meaning that no stars will have yet begun nuclear burning. This however would affect our lower resolution simulations, which run for significantly longer periods, although the sink particles in those simulations (potentially) represent groups of protostars rather than individual stars. A comprehensive treatment of protostellar feedback is nonetheless outside the scope of this study.

Lastly, as protostellar mergers are often reported in primordial star forming simulations (e.g. \citealt{Greif2012,Hirano2017,Susa2019}), we include the treatment of sink particle mergers originally implemented in \cite{Prole2022}. Sink particles are merged if:
   \begin{enumerate}
      \item They lie within each other’s accretion radius.
      \item They are moving towards each other.
      \item Their relative accelerations are <0.
      \item They are gravitationally bound to each other.
   \end{enumerate}
Since sink particles carry no thermal data, the last criteria simply require
that their gravitational potential exceeds the kinetic energy of
the system. When these criteria are met, the larger of the sinks gains
the mass and linear momentum of a smaller sink, and its position
is shifted to the centre of mass of the system. We allow multiple mergers per time-step. For example, if
sink A is flagged to merge into sink B, and sink B is flagged to
merge into sink C, then both A and B will be merged into sink C
simultaneously.


\section{Initial collapse}
\label{sec:collapse}
Figure \ref{fig:sink_formation} summarises the state of the collapse at a point just before the formation of the first sink particle. The temperature-density relationship in the top panel shows that the initial contraction up to densities of $10^{-15}$ g cm$^{-3}$ follows a near isothermal  collapse as expected (e.g. \citealt{Omukai2000,Klessen2023}). 
Above this density, the temperature drop steepens, but remains close to isothermal. Here the temperature drops by around a factor of 2 over a density range of $10^4$, giving an effective gamma of $\sim 0.95$. This steepening of the temperature profile was first reported on in the one-zone calculations of \citet{Omukai2001b}; at this density, the H$_2$ abundance is still too low to provide significant cooling, instead the temperature drop corresponds to the point where cooling becomes dominated by H$^{-}$ free-bound cooling. 
The near-isothermal collapse of the gas is further evidenced in the bottom panel of Figure \ref{fig:sink_formation} where we clearly see a $\rho \propto R^{-2}$ scaling of the density versus radius over more than six orders of magnitude in radius.

From the very bottom panel of Figure \ref{fig:sink_formation}, the H$_2$ abundance begins to build up, albeit from very low abundance levels, within the inner 10$^{-2}$ pc ($\sim$ 2000 au) of the halo. While our LW field strength of $J_{21}=10^5$ is already quite extreme, we run a second realisation of Halo 3 with an extremely high value of $10^{10}$ (as shown in the top panel of Figure \ref{fig:shielding}) to demonstrate that above roughly 10$^{-15}$ g cm$^{-3}$ LW photo-dissociation can not prevent an increase in the H$_2$ abundance (nor probably can any other physical process). The three-body H$_2$ formation rate per unit volume is proportional to $n^3$ (where $n$ is number density), whereas the corresponding scalings for the photodissociation rate and the H$_2$ collisional
dissociation rate are $n$ and $n^2$, respectively. Therefore H$_2$ formation will inevitably overcome its destruction at high densities. In practice then the formation of \molH at high densities is inevitable. 

Furthermore, a build in the H$_2$ fraction at high densities owes to the exponential nature of the self-shielding. To demonstrate this, we calculate the H$_2$ shielding factor given in \cite{Wolcott-Green2011} as
\begin{equation}
 f_{\rm H_2} = \frac{1}{(1+\frac{N_{ \rm H_2}}{2.34 \times 10^{19}})^{2.38 \times 10^{-1}}} {\rm exp} \left( \frac{-5.2 \times 10 ^{-3} N_{\rm H_2}}{2.34 \times 10^{19}} \right),
\end{equation}
where $N_{ \rm H_2}$ is the H$_2$ column density. The shielding factor acts as a transmission factor for the LW radiation, with lower values representing more effective shielding. We calculate values of $N_{ \rm H_2}$ by extrapolating from the column density-number density power law relation shown in Figure 3 of \cite{Wolcott-Green2011}. The shielding factor is given as a function of density in the bottom panel of Figure \ref{fig:shielding}, which quickly falls to 0 above densities of 10$^{-15}$ g cm$^{-3}$, independently of the LW intensity. This shows that the core will always be shielded from LW radiation above these densities regardless of the value of $J_{21}$.
A build up of H$_2$ fraction within the dense core has been seen to various degrees in the radial profiles of previous studies. For example, the one-zone calculations of \citet{Omukai2001b} show a \molH fraction of $\sim 0.01$ at a density of $\sim 10^{-8}$ g cm$^{-3}$, while simulations by \citet{Becerra2015} reach a \molH fraction of $\sim 0.1$ by $\sim 10^{-6}$ g cm$^{-3}$ and \citet{Regan2017} reaches a \molH fraction of 10$^{-4}$ by their maximum density of $10^{-15}$ g cm$^{-3}$. What the results here show is that the very central regions of atomic cooling halos are highly likely to have small pockets of fully molecular gas.


\begin{figure}
	 \hbox{\hspace{-1cm} \includegraphics[scale=0.73]{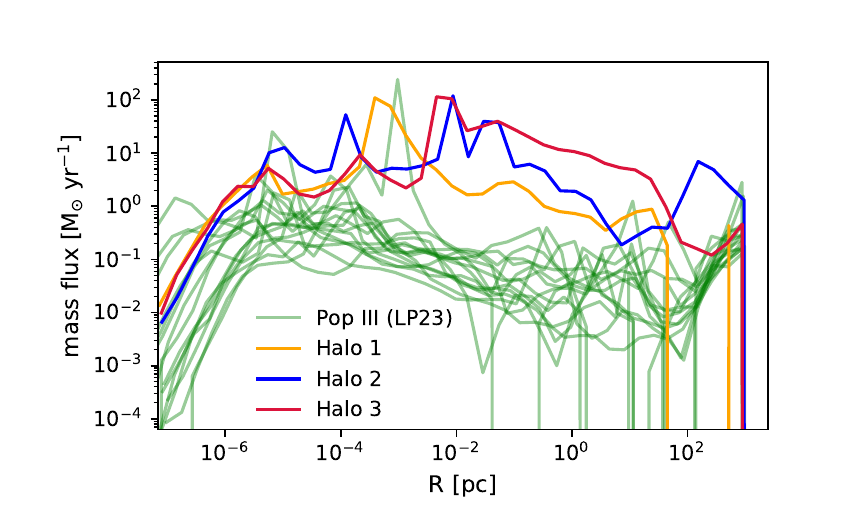}}
    \caption{Radial profile of the mass flux through consecutive shells i.e. the accretion rate onto the center of the halo. For comparison, results from the 15 Pop III star forming minihalos from \citetalias{Prole2023} are shown in green.}
    \label{fig:accretion}
\end{figure}

\begin{figure}
	 \hbox{\hspace{-0.8cm} \includegraphics[scale=0.8]{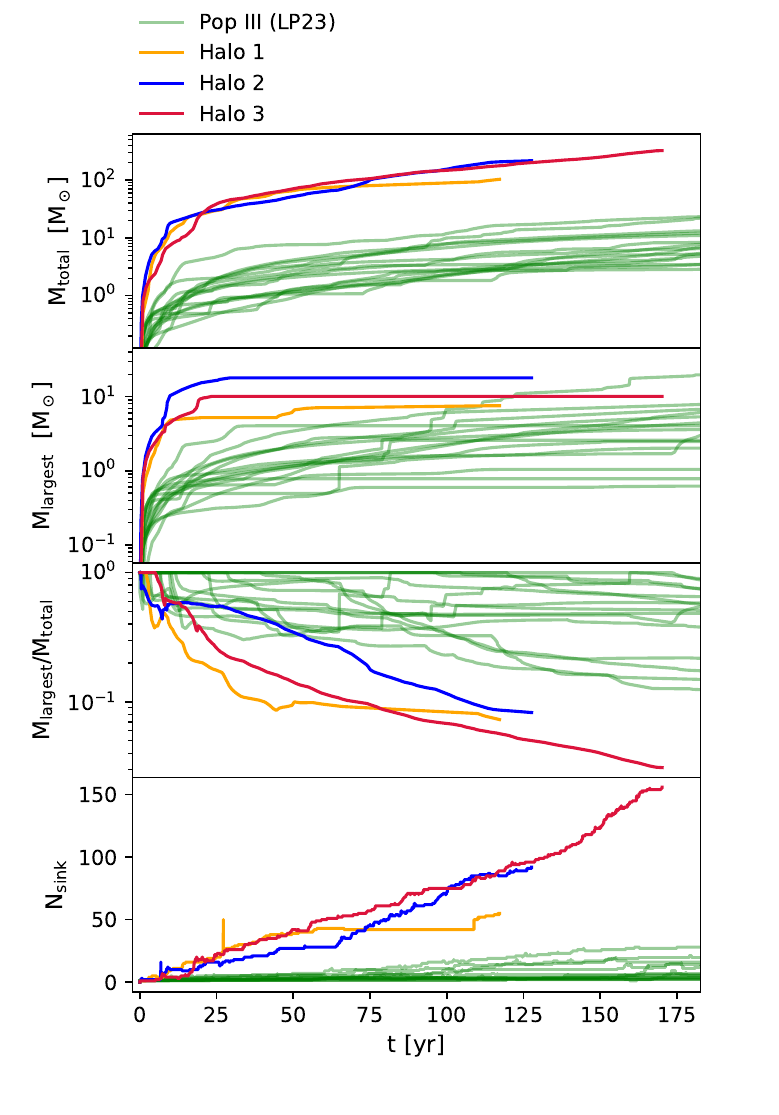}}
    \caption{Time evolution of the ratio of mass of the largest sink to total mass accreted accross all sinks, mass of the largest sink particle, total mass accreted onto sink particles and total number of sink particles. For comparison, results from the 15 Pop III star forming minihalos from \citetalias{Prole2023} are shown in green. }
    \label{fig:fragmentation}
\end{figure}

\begin{figure}
	 \hbox{\hspace{-0.7cm} \includegraphics[scale=0.75]{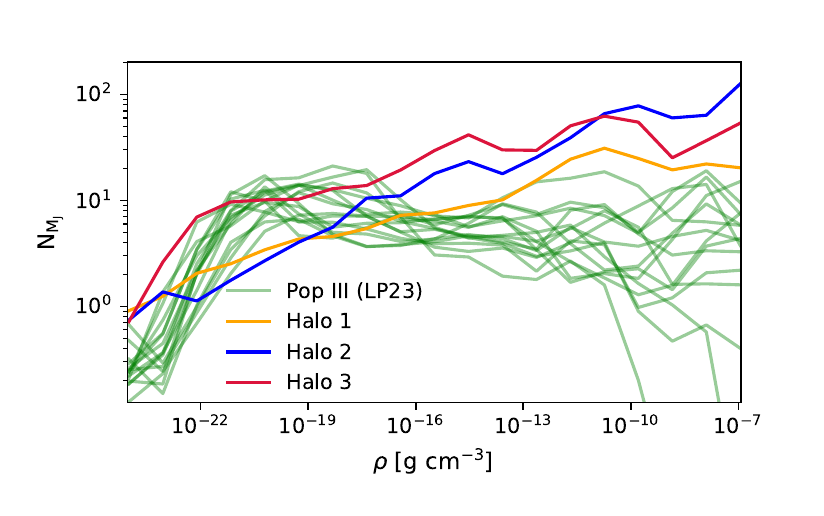}}
    \caption{Number of enclosed Jeans masses as a function of density, shown at a time 100 yr after the formation of the first sink particle. The Jeans mass at each density bin is calculated using the mass weighted average temperature and density within the bin. The calculated Jeans mass is then compared to the total mass of gas at or above the density of the bin.}
    \label{fig:jeans}
\end{figure}

\section{Fragmentation and the Initial Mass Function}
\label{sec:frag}
We now move onto to discuss the build of the initial mass function and the resulting protostellar masses that are found in our simulations at the very highest densities when we introduce sink particles. In Figure \ref{fig:accretion} we show the mass flux into concentric shells just before the formation of the first sink particle, comparing the standard Pop III star forming minihalos ($\sim 10^6$ M$_\odot$) of \citetalias{Prole2023} with the more massive, atomically cooled halos simulated here.  Within the inner $\sim$100 pc, the accretion rate into the atomic halos exceeds the minihalo case by a factor of between 10-1000 due to the combined effects of a stronger DM gravitational well and a larger available reservoir of baryonic gas. In principle this should lead to the formation of more massive sinks given the higher accretion rates possible.

Figure \ref{fig:fragmentation} shows how the total mass in sink particles, the mass of the most massive sink particle and the total number of sinks evolves in time. The total mass in sinks of the system exceeds the minihalo case by almost 2 order of magnitude due to the higher mass flux. While the initial accretion onto the most massive sink particle far exceeds the same quantity in the minihalos, the most massive sink is ejected from the system in all three atomic halos within the first $\sim$50 yr after its formation. This results in the most massive sink particle aligning with the upper limit seen in the minihalos. The most surprising result is that the fragmentation of the gas in the center of the atomic halos far exceeds the minihalo cases. 

As discussed earlier, H$^-$ formation becomes the dominant cooling process above 10$^{-15}$ g cm$^{-3}$, leading to a slightly steeper decline in temperature with density. However, the temperature is still higher in the atomically cooled halos when compared to the minihalo case at the same density (see e.g. \citealt{Omukai2000,Yoshida2006,Prole2022}), corresponding to a higher Jeans mass. The increase in fragmentation is therefore attributable to the higher mass infall rate. For example, assuming fragmentation is perfectly efficient, the maximum number of fragments that can form is the number of Jeans masses present within the disc. We can therefore get an estimate of the fragmentation from Figure \ref{fig:jeans}, which shows the number of enclosed Jeans masses of gas at or above a given density. At high densities the number of Jeans masses present in the atomically cooled halos exceeds the minihalo case by an order of magnitude and roughly matches the number of sink particles formed in each halo (see Figure \ref{fig:fragmentation}), which explains the increase in fragmentation when compared to the minihalos.

\begin{figure}
	 \hbox{\hspace{-0.6cm} \includegraphics[scale=0.72]{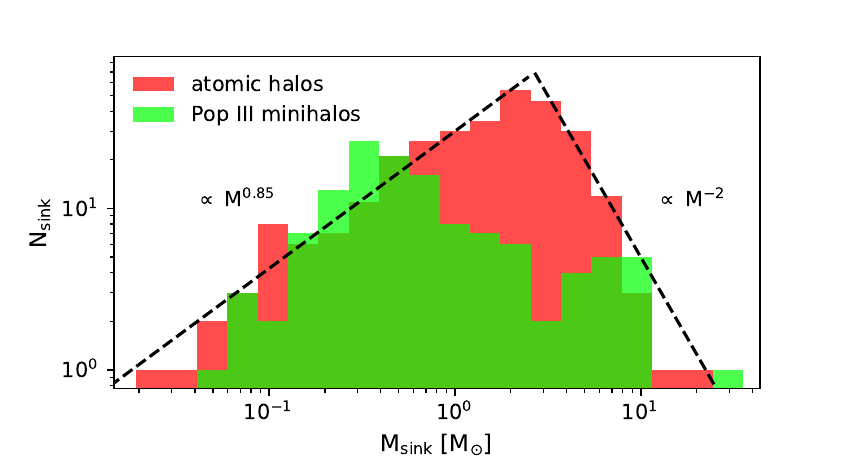}}
    \caption{Comparison of the sink particle mass function from the 3 atomic cooling halos at $\sim$100 yr versus the 15 H$_2$ cooling minihalos from \citetalias{Prole2023} at $\sim$300 yr. Power laws of M$^{0.85}$ and M$^{-2}$ are superimposed to give the reader an idea of the slopes involved.}
    \label{fig:IMF}
\end{figure}

The combined sink particle mass functions from all three halos are shown in Figure \ref{fig:IMF} at $\sim100$ yr after the formation of the first sink particle. We also show the mass functions from the minihalos at the end of the \citetalias{Prole2023} simulations ($\sim$300 yr). Despite increased fragmentation and a factor of 3 less in time to accrete, the atomic halo mass function peaks at a higher protostellar mass of 3 M$_\odot$ compared to the 0.3 M$_\odot$ peak in the minihalos. We note that these protostars will continue to accrete for roughly 10$^4$ yr before the end of the pre-main sequence. While the zero-age main sequence (ZAMS) masses of these stars are unknown (see \S \ref{sec:mass}), we have confirmed that atomically cooled halos do produce a population of higher mass protostars when compared to regular Pop III star forming minihalos, at least after the first 100 yr after the formation of the first protostar. These protostars can then accrete the available gas in competition with further star formation.

\begin{figure}
	 \hbox{\hspace{-0.8cm} \includegraphics[scale=0.73]{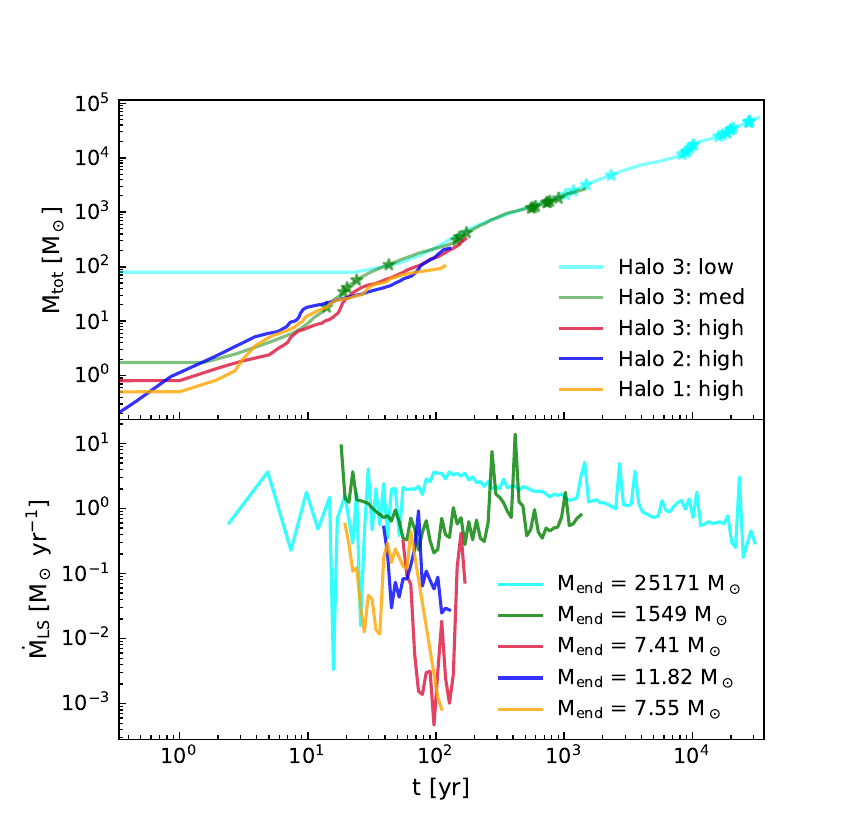}}
    \caption{Comparison between the high resolution and low resolution simulations. Top - time evolution of the total mass of the system. The times at which new sink particles formed are indicated with star-shaped markers in the medium and low resolution simulations. We do not show the formation of sink particles for the high resolution simulations as fragmentation occurs almost instantly. Bottom - accretion rates onto the most massive surviving (non-ejected) sink particle. Also shown are the masses of the most massive surviving sink at the end of the simulations.}
    \label{fig:lowres}
\end{figure}

\section{Final stellar and black hole masses} 
\label{sec:mass}
In order to estimate the ZAMS masses of the sinks we run lower resolution simulations of Halo 3 which allows us to evolve the system over longer timescales.
The top panel of Figure \ref{fig:lowres} shows how the total mass of the system of sink particles grows as a function of time across the different resolutions tested. We show the formation of new sink particles in the 10$^{-13}$ and 10$^{-10}$ g cm$^{-3}$ simulations as star shaped markers. We do not show the formation of sinks for the 10$^{-6}$ g cm$^{-3}$ cases as fragmentation occurs almost immediately. Clearly the higher the resolution used, the earlier fragmentation occurs. If we take $t = 0$ to be the time at which the first sink forms in each case then the second sink forms almost immediately in the highest resolution case, after $\sim 10$~yr in the $10^{-10} \: {\rm g \, cm^{-3}}$ case and only after $\sim 1000$~yr in the $10^{-13} \: {\rm g \, cm^{-3}}$ case. As the resolution used does not significantly affect the growth of the total system, the lowest resolution run shows that the system will continue to grow linearly through the pre-main sequence phase to reach a mass of $\sim10^4$ M$_\odot$ by 10$^4$ yr i.e. there will be $\sim10^4$ M$_\odot$ available for star formation within the central 264 au. The increased fragmentation in the higher resolution simulations complicates how this mass will be distributed amongst the protostars. The bottom panel of Figure \ref{fig:lowres} shows the accretion rate onto the most massive surviving sink particle i.e. the most massive non-ejected protostar. At the end of the high resolution simulations, the largest survivors in the three halos have accretion rates in the range 10$^{-3}$-10$^{-1}$ M$_\odot$ yr$^{-1}$ and have masses in the range of 7.5-12 M$_\odot$. Among the two lower resolution simulations the final masses of the 
protostars are approximately 1500 M$_\odot$ (after 1000 years) and 25000 M$_\odot$ (after 20,000 years) respectively. If the most massive sink particle from the highest resolution simulations can accrete or grow at or near the rates found in the lower resolution simulations then super-massive star formation may be realised.

The zero-age main sequence (ZAMS) mass of a star depends on the pathway between the formation of the protostar and the eventual beginning of core hydrogen burning, which in turn depends strongly on the evolution of the protostellar accretion rate. When a protostar's Kelvin–Helmholtz (KH) timescale (time to radiate away its own gravitational energy) is shorter than its accretion timescale (time to double its mass by accretion), it will radiate away its energy and begin to contract, which typically occurs at masses $\sim$10 M$_\odot$ \citep{Palla1991,Omukai2003} though this depends sensitively on the assumed accretion rate \citep[e.g.][]{Nandal2023}. The contraction causes an increased extreme ultraviolet (EUV) luminosity and surface temperature, ionizing infalling gas and accelerating it outwards. This triggers a runaway collapse as the decreased accretion rate causes further contraction until hydrogen burning begins and the protostar reaches the ZAMS. If the accretion rate onto the largest sink particle in our high resolution simulations falls and remains below $\sim 4 \times 10^{-3}$ M$_\odot$ yr$^{-1}$, the KH contraction will begin at $\sim$10 M$_\odot$ and the growth of the protostar will be self-regulated by these radiative feedback effects, limiting the final mass to a few tens of M$_\odot$ \citep{Hosokawa2011a,Hosokawa2012}. However, numerous studies have examined the impact that accretion has on the contraction of a Pop III star \citep[e.g.][]{Hirano2014}. A key quantity here is the critical accretion rate - this is the accretion rate onto the stellar surface that prevents contraction of the star. Recently \cite{Nandal2023} investigated in detail this quantity in terms of episodic accretion rates using the Geneva Stellar Evolution code ({\sc Genec}: \citealt{Eggenberger2008}) . They found that during the pre-main sequence, which is of most relevance here, the critical accretion rate is $\sim 2.5 \times 10^{-2}$ M$_\odot$ yr$^{-1}$. In this case the effective surface temperature remains below $10^4$ K, hence feedback can not form a H\MakeUppercase{\romannumeral 1}\MakeUppercase{\romannumeral 1} region and accretion is not prevented, leading to the formation of a super-giant protostar, which can grow up to several thousand solar masses depending on the details of the accretion \citep{Umeda2016,Woods2017,Haemmerle2018}. The end of accretion onto a super-giant protostar is caused by a fast contraction when the accretion rate falls below $\sim 7 \times 10^{-3}$ M$_\odot$ yr$^{-1}$.


At the end of the high resolution simulations, the accretion rates onto the largest surviving sink particles vary significantly in time. As the accretion rates onto $\sim$10 M$_\odot$ protostars appear to be a good indicator of future contraction/accretion evolution \citep{Hosokawa2011a,Hosokawa2012,Hirano2017} and the largest survivor in Halos 2 and 3 have accretion rates between 10$^{-2}$-10$^{-1}$ M$_\odot$ yr$^{-1}$, they could feasibly go on to form (super-massive) Pop III stars with masses anywhere in the range of 10 M$_\odot$ up to $10^4$ M$_\odot$ within their main sequence (MS) lifetime. If they grow in excess of 25 M$_\odot$ the type II supernova explosion would be too weak to eject much of the star  and the subsequent fallback of material causes the resulting neutron star to collapse into a BH \citep{MacFadyen2001}, while above 40 M$_\odot$ the neutron star is unable to launch a supernova shock and collapses directly to form a BH with no mass loss \citep{Fryer1999}. If these stars avoid the disruptive pair instability supernova mass range of 140-260 M$_\odot$, they will result in a massive BH equal in mass to the stellar progenitor. However, whether the protostars will maintain their accretion rates is uncertain. In the optimistic case that they maintain accretion rates of 10$^{-1}$ M$_\odot$ yr$^{-1}$, these results are in line with many previous, lower resolution simulations of heavy seed black hole formation (e.g. \citealt{Johnson2011c,Latif2013c,Latif2013b,Regan2014,Latif2015a,Choi2015,Regan2017,Smidt2017,Ardaneh2018}) despite the increased gas fragmentation. The largest survivor in Halo 1 ends with an accretion rate of $\sim$10$^{-3}$ M$_\odot$ yr$^{-1}$, which if maintained will likely lead to an early KH contraction and limit the final mass to 10-30 M$_\odot$. It is unclear if the accretion rate onto Halo 1 will remain below 10$^{-3}$ or if it will experience an increase similar to that of Halo 3. Certainly the formation of massive BH seeds of 10$^{4}$ M$_\odot$ needed to explain $z\sim 7$ quasar observations would rely on frequent mergers with other protostars. 

The sink particle masses in the lower resolution simulations represent whole groups of protostars in the high resolution simulations. Frequent mergers of secondary protostars into the main protostar would push the stellar and later BH masses up to close to what is achieved in the low resolution simulations, with an upper limit of 10$^4$ M$_\odot$ within a 10 kyr period. Recent studies have pointed out the importance of super competitive accretion, in which a central few stars grow supermassive while a large number of other stars are competing for the gas reservoir \citep{Chon2020}, with central objects growing significantly through mergers \citep{Sassano2021,Vergara2021,Trinca2022,Schleicher2022,Zwick2023}. Here the mass growth by collisions can be comparable to its growth via accretion \citep{Schleicher2023}. The super-competitive accretion scenario was initially invoked to explore the 'low metallicity' regime ($\rm{Z} \sim 10^{-6} - 10^{-3}$ Z$_{\odot}$) where fragmentation is expected to be more active compared to the metal-free case. What we find here is that fragmentation is already vigorous within the central 2000 au of the halo. It therefore appears that for all metallicities below $\sim 10^{-3}$ Z$_{\odot}$ we can expect a scenario where vigorous fragmentation competes with a rapidly growing protostar. Differentiating the environments which produce a dense stellar cluster versus those that produce a central massive Pop III star may be the next frontier. 

\begin{figure}
	 \hbox{\hspace{-0.6cm} \includegraphics[scale=0.75]{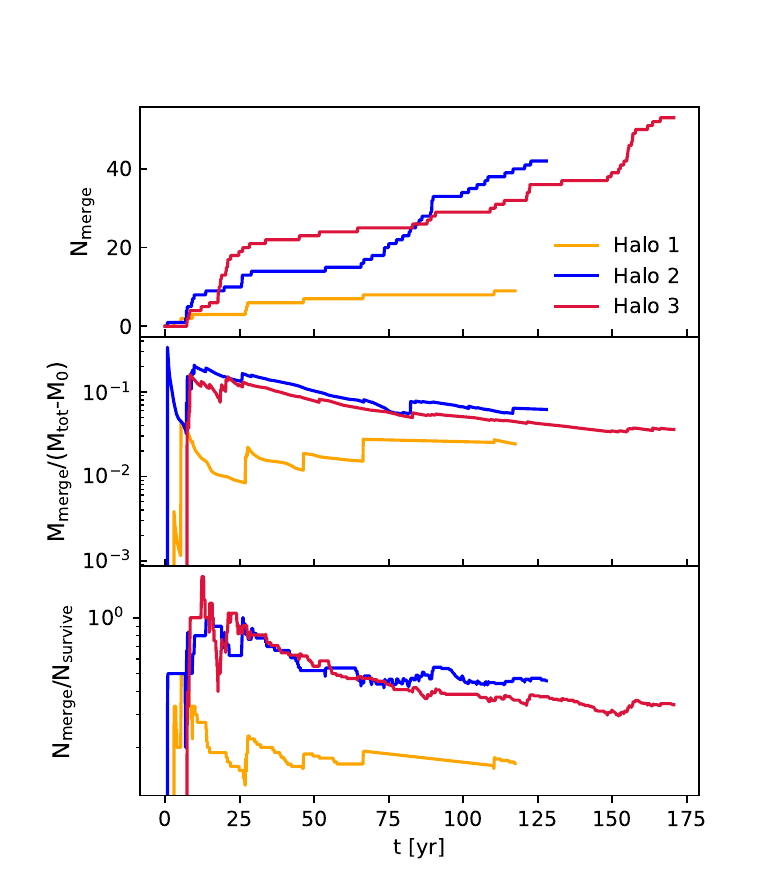}}
    \caption{Time evolution of the merger data. Top - cumulative number of sink particle mergers. Middle -  ratio of cumulative merged mass to the total accreted sink particle mass (M$_{\rm tot}-$M$_{0}$) where M$_{0}$ is the initial mass of a sink particle. Bottom - ratio of cumulative number of sink mergers against the current number of surviving sink particles.}
    \label{fig:mergers}
\end{figure}

To investigate the merger behaviour further, we plot the total number of sink mergers, ratio of total merged mass to accreted sink mass and the ratio of the number of mergers to the number of surviving sink particles as a function of time in Figure \ref{fig:mergers}. From this we see that the mass gained through mergers constitutes between 1-10$\%$ of the total sink particle mass. The frequency of mergers is high, with the number of mergers totalling between 10-50$\%$ of the total number of surviving sink particles. If this holds throughout the following 10$^4$ yr, the total mass gained through mergers alone could be as high as 10$^3$ M$_\odot$, although it is unclear if this mass would be shared between the growing number of protostars randomly or be preferentially received by the largest growing BH seed.

\section{Caveats} 
\label{sec:caveats}

We have not included magnetic fields in these simulations. While studies of primordial magnetic fields suggest that they can increase the mass of protostars (e.g. \citealt{Saad2022,Hirano2022,Stacy2022}), the fields have no effect when they are properly resolved, distributing the magnetic energy from the small-scale turbulent dynamo to smaller spatial scales \citep{Prole2022a}. 

We have also not included radiative feedback from our protostars, which has the effect of heating the surrounding gas and lowering accretion rates. While our high resolution simulations end at a point when the protostars would not produce significant levels of feedback. The low resolution simulations however run for much longer and would be subject to feedback effects.

As our high resolution simulations have only captured the first $\sim 100$ yr of accretion after the formation of the first protostar and the lower resolution simulations lack the capacity to resolve individual star foratmion, to what extent competitive accretion and mergers affect the subsequent final stellar/BH masses is unclear.

Ideally we would resolve up to the protostellar formation density of 10$^{-4}$ g cm$^{-3}$, which is currently computationally unfeasible. Failure to resolve up to that density means we have not achieved numerical convergence. Despite this, our implemented threshold density of 10$^{-6}$ g cm$^{-3}$ and post-sink particle run time of 100 yr represents the current state of the art in this regime.

As mentioned in \S \ref{sec:chem}, although we account self-consistently for the effects of opacity when computing the H$_2$ cooling rate (due to line cooling and collision-induced emission), we do not yet account for the continuum opacity of the gas when computing the cooling rate due to other radiative processes, e.g.\ H$^{-}$ free-bound cooling. At most densities in our models, the optical depth of the gas in the continuum is small and this simplification has little impact -- for example, it should be valid at all of the densities traced in our medium and low resolution models. However, above a density of $\sim 10^{-8} \: {\rm g \: cm^{-3}}$, we expect the optical depth of the gas in the continuum to become significant, and hence our current simulations over-estimate the cooling rate of the gas above this density. This will likely give us slightly more effective fragmentation at the highest densities than we would find in reality. However, we do not expect this to significantly impact the main conclusions of our study. As \ref{fig:jeans} demonstrates, the number of Jeans masses present in the gas at $\rho = 10^{-8} \: {\rm g \: cm^{-3}}$ in the atomic cooling halos already substantially exceeds the corresponding value in H$_{2}$-cooled minihalos, demonstrating that this result is not a consequence of our simplified treatment of very high density cooling. Our finding that the gas fragments extensively in the centre of the atomic-cooled halos should therefore be a robust result, although some of the details (e.g. the precise number of fragments and their formation masses) will depend on the treatment of the cooling at very high densities. Our finding that the fragments quickly grow to become more massive than their Pop III counterparts should also be a robust result: properly accounting for the continuum opacity will likely give us fragments with slightly larger initial masses, which if anything will increase the rate at which they later gain mass, exacerbating the difference between the atomic-cooled and H$_{2}$-cooled results.

We have assumed that the background radiation field from Pop III stars takes the form of a black body spectrum with an effective temperature of 10$^5$ K, peaking in the UV. However, this relies on the stars having masses of $\sim$ 300 M$_\odot$ \citep{Schaerer2002}. The masses of Pop III stars are currently uncertain, so our choice of effective temperature affects the photodissociation  of the molecules present in the gas. For example, the effective temperature for a 1 M$_\odot$ Pop III star is $\sim$ 7000 K \citep{Larkin2022}, which peaks at lower frequencies, resulting in significantly less H$_2$ photodissociating LW radiation but more infrared radiation, which can still photodissociate H$^-$ to disrupt H$_2$ formation \citep{Chuzhoy2007}.

The methodology here was idealised in that we used  a six species chemical model to construct a pristine atomic cooling halo before switching to a full 12 species chemical model and employing an intense LW background. In future work we will apply the same methodology to rapidly assembling halos without the reliance on super-critical LW values.

\section{Conclusions}
\label{sec:conclusions}
We have performed high resolution simulations of three pristine atomically cooled halos of mass $\sim10^8$ M$_\odot$ which begin to collapse at $z\sim 11.5-13.3$. The goal of our simulation suite was to examine the fragmentation and protostellar formation within the core of such systems as they reach protostellar densities. For the highest resolution we could tractably simulate ($\rho \sim 10^{-6}$ g cm$^{-3}$) we followed the protostar formation for approximately 100 years by introducing sink particles representing individual protostars. The main results are as follows:
   \begin{enumerate}
      \item Resolving protstellar densities reveals that the center of (pristine) atomically cooled halos is subject to intense fragmentation, even more-so than in the canonical Pop III minihalo case. This is driven by H$^{-}$ free-bound cooling and an increased number of Jeans masses within the core, owing to enhanced accretion rates.
      \item Despite increased fragmentation, the atomically cooled halos formed a population of higher mass protostars compared to Pop III star forming minihalos.
      \item The initial accretion rates onto the most massive surviving protostars indicates that their zero-age main sequence masses could range from $10^2 - 10^4$ M$_\odot$ depending on subsequent accretion and/or mergers. 
      \item Our coarser resolution simulations show that the total mass of the system continues to grow steadily for at least 10$^4$ years after the formation of the first protostar (achieving central sink particle masses in excess of $10^4$ \msolarc), although how this is distributed amongst the rapidly growing number of protostars is unknown and requires further study.
      \item The formation of a massive Pop III star (M$_* \gtrsim 1000$ \msolarc) is therefore realistic and achievable in a high-z setting but relies on (super-competitive) accretion and frequent mergers with secondary protostars
      \item The H$_2$ fraction within the inner $\sim$2000 au of all three halos was able to build up independently of the strength of the external LW radiation field. This strongly indicates that all collapsing halos (even those subject to strong LW fields and likely other physical processes which could suppress star formation up to the atomic limit e.g. streaming velocities between DM and baryonic gas) contain small pockets of cold, H$_2$ rich gas at their centre.
   \end{enumerate}

\noindent The findings here show that the formation of massive Pop III stars (i.e. heavy seed MBHs) is entirely plausible in sufficiently massive halos but will depend sensitively on the future evolution of the protostars and on the balance between stellar accretion and mergers in the dense cluster that forms within the center of the atomic cooling halo. While we model the impact of a super-critical LW radiation field here, any physical process which results in the initial suppression of star formation up to the atomic cooling limit should have a similar impact and allow for the formation of more massive Pop III stars.
\begin{acknowledgements}
LP and JR acknowledge support from the Irish Research Council Laureate programme under grant number IRCLA/2022/1165. JR also acknowledges support from the Royal Society and Science Foundation Ireland under grant number URF\textbackslash R1\textbackslash 191132. 

\ \
This work used the DiRAC@Durham facility managed by the Institute for Computational Cosmology on behalf of the STFC DiRAC HPC Facility (www.dirac.ac.uk). The equipment was funded by BEIS capital funding via STFC capital grants ST/P002293/1, ST/R002371/1 and ST/S002502/1, Durham University and STFC operations grant ST/R000832/1. DiRAC is part of the National e-Infrastructure.

\ \ 
We also acknowledge the support of the Supercomputing Wales project, which is part-funded by the European Regional Development Fund (ERDF) via Welsh Government.

\ \
SCOG and RSK acknowledge financial support from the European Research Council via the ERC Synergy Grant “ECOGAL” (project ID 855130), from the Heidelberg Cluster of Excellence (EXC 2181 – 390900948) “STRUCTURES”, funded by the German Excellence Strategy, and by the German Ministry for Economic Affairs and Climate Action in project ``MAINN'' (funding ID 50OO2206). They also thank for computing resources provided by the Ministry of Science, Research and the Arts (MWK) of the State of Baden-W\"{u}rttemberg and Deutsche Forschungsgemeinschaft (DFG) through grant INST 35/1134-1 FUGG and for data storage at SDS@hd through grant INST 35/1314-1 FUGG.

\ \ 
Finally, we acknowledge Advanced Research Computing at Cardiff (ARCCA) for providing resources for the project.
\end{acknowledgements}

\bibliographystyle{mnras}
\bibliography{references.bib}

\end{document}